\documentstyle[12pt,aasms4]{article} 
\slugcomment{version : astro-ph jul 4,2000}

\newcommand{\beq}{\begin{equation}}
\newcommand{\eeq}{\end{equation}}
\def\etal{{\it et~al.\ }}
\def\eg{{\it e.~g.\ }}

\newcommand{\hii}{H~{\sc ii}~}
\lefthead{Franco \etal}
\righthead{Decreasing Density Gradients in the Circumnuclear \hii Regions of Barred Galaxies} \normalsize
\begin{document} 
 
\title{Decreasing Density Gradients in Circumnuclear \hii Regions of Barred Galaxies NGC 1022, 1326 \& 4314}

\author{Jos\'e Franco\altaffilmark{1}, J. Antonio Garc\'{\i}a-Barreto\altaffilmark{1} and Eduardo de la Fuente\altaffilmark{1,2}}

\altaffiltext{1}{Instituto de Astronom\'{\i}a, Universidad Nacional Aut\'onoma 
de M\'exico, Apartado Postal 70-264, M\'exico D.F. C.P. 04510}

\altaffiltext{2}{Facultad de Ciencias, Universidad Nacional Aut\'onoma de
M\'exico, M\'exico D.F. C.P. 04510}

\begin{abstract}

A re-analysis of the radio continuum emission from circumnuclear regions of the
barred galaxies NGC 1022, 1326 \& 4314 is presented. The spatial distributions 
of H$\alpha$ and radio continuum are similar and suggest the existence of 
several giant \hii regions in the circumnuclear zone. The spectral indices of 
the \hii regions, obtained with similar angular resolution and at three 
different wavelengths, indicate that in all three galaxies $\alpha^6_2 >-0.1$. 
This result suggests the existence of giant \hii regions with decreasing
density gradients. The average density stratifications of these \hii
regions could be approximated by power-laws of the form $n_e\propto
r^{-\omega}$, with exponents in the range $1.6 < \omega < 2.4$. 

\end{abstract}


\section{Introduction}

The average density profiles of extragalactic \hii regions can be derived from 
their thermal radio emission. The thermal 
radio continuum emission of optically thin photoionized plasma with $\tau_{ff} 
<1$ has a relatively flat spectrum with S$_{\nu}~{\propto}~\nu^{-0.1}$, and it 
changes to S$_{\nu}~{\propto}~\nu^2$ when the region becomes optically thick,
with $\tau_{ff}>1$, and has a sharp boundary (\eg Mezger, Schraml \& Terzian 1967;
Mezger \& Henderson 1967). When the electron density varies with distance, say 
as a power-law with $n_e~{\propto}~r^{-\omega}$, and the plasma is optically 
thick near the center of the emitting object (\eg ionized stellar winds or dense
 photoionized molecular clouds), the spectral index varies as S$_{\nu} \propto 
\nu^{\alpha}$, with $-0.1<{\alpha}<+2$ for $\omega$ ranging from 0 to $\infty$ 
(Olnon 1975; Panagia \& Felli 1975). This behavior is indeed observed in stars
with extended winds (\eg Simon \etal 1983) and in some galactic ultra-compact
and super-ultra-compact \hii regions (Franco, Kurtz \& Hofner 2000). In
particular, the radio continuum emission of the core of galactic source G35.20--1.74 scales as
S$_{\nu} \propto \nu^{0.6}$, indicating that its internal density structure is
proportional to $r^{-2}$ (Kurtz 2000; Franco \etal 2000).
 
Here, evidence is given that the average density profile of extragalactic \hii
regions can indeed be derived from the radio continuum, and discuss the radio
emission from the circumnuclear regions in three barred galaxies: NGC 1022,
1326 and 4314. Their H$\alpha$ and radio continuum emission at 20, 6, and
2 cm, along with CO, HI, and near IR have been reported by Garc\'{\i}a-Barreto
\etal (1991a,b,c, 1996). In this paper we re-analyze the radio continuum data
and report the spectral indices of the radio continuum emission from several
sources in the circumnuclear regions of these galaxies. Their spectral indices
between 20, 6, and 2 cm, with observations at the same angular
resolution, indicate contributions from both synchrotron and free-free
emission. Our results indicate: 1) the existence of several circumnuclear \hii
regions with radio continuum spectral indices $\alpha^6_2>-0.1$, and 2) the
spectral index between 6 and 2 cm from each of the circumnuclear \hii
regions can be explained if the photoionized gas is optically thick and has a
decreasing density structure with a power-law form $n_e\propto r^{-\omega}$,
where the exponent is in the range $1.6<{\omega}<2.4$.
 
\section{The Ionized Gas in NGC~1022, NGC 1326, and NGC 4314}

NGC~1022 is an SBa(r)p barred spiral galaxy located in the Cetus-Aries group of
galaxies, at a  distance of 18.5 Mpc (H$_0=75$ km s$^{-1}$ Mpc$^{-1}$; Tully 
1988). The radio continuum emission at 20, 6, and 2 cm (with the same 
angular resolutions), CO (1-0) and CO (2-1) emission, and near IR with a single
element detector have been reported by Garc\'{\i}a-Barreto \etal (1991c). NGC 
1326 is an RSBa barred spiral galaxy in the Fornax I cluster of galaxies  at a 
distance  of 16.9 Mpc (Tully 1988). Radio continuum emission at 20 cm, 6 cm,
and 2 cm, CO (1-0) and CO (2-1) emission, and optical spectroscopy have been
reported by Garc\'{\i}a-Barreto \etal (1991b). NGC 4314 is an SB(a) barred
spiral galaxy in the Coma I group of galaxies, at an adopted distance of 10
Mpc. Radio continuum emission at 20 cm, 6 cm, and 2cm, HI, CO (1-0) and CO
(2-1) emission, and near IR with a single element detector have been reported
by Garc\'{\i}a-Barreto \etal (1991a). Images of the optical continuum in the I
filter (8040 \AA) and continuum-free H$\alpha$+[NII] of these three
galaxies have been reported by Garc\'{\i}a-Barreto \etal (1996). 

Here we will focus on the properties of the radio continuum emission of the
different sources in the circumnuclear regions. In all three galaxies the
H$\alpha$+[NII] and the radio continuum coincide spatially, indicating that the
emission is associated with the star formation process. The radio continuum
emission maps were obtained at the VLA with similar high angular resolution
beams at 20 cm, 6 cm and 2 cm in order to do spectral index analysis of similar
regions at the three wavelengths (Garc\'{\i}a-Barreto \etal 1991a,b,c). The
H$\alpha$ line emission images were obtained with a CCD camera at the 2.1m
telescope in San Pedro M\'artir, B.C. M\'exico (Garc\'{\i}a-Barreto \etal
1996). 

Figure 1 is a superposition of the radio continuum emission at 2 cm 
on the H$\alpha$ emission of NGC 1022. Both the radio
continuum and H$\alpha$ emitting regions are distributed in a similar manner: 
there is a bright emission region coincident with the compact nucleus, and 
there are two additional compact sources off-nucleus (to the north-east and 
slightly north-west). We have labeled them as NGC 1022:FGF 1 (nucleus), 2 (to the northeast), and 3 (to the northwest), respectively (see Figure 1 and Table 1). In
the original analysis, done in 1991, only the two strongest sources, 1 and 2,
were considered in the radio continuum analysis (Garc\'{\i}a-Barreto \etal
1991c); however, analysis of the H$\alpha$ distribution suggested weak emission
from a source to the NW (see Fig. 1 and Garc\'{\i}a-Barreto \etal 1996), and a
closer inspection of the radio continuum maps also indicated the existence of 
source 3 at the same location. At larger scales, Figure 2 shows that the
radio continuum emission is found only in the central regions of the disk. The
image shows the optical I continuum superimposed on the radio
continuum at 20 cm. Notice that the position angle of the
continuum emission between source 1 (the nucleus) and source 2 is almost
perpendicular to the position angle of the stellar bar (P.A.$_{radio} \sim
25^{\circ}$ vs. P.A.$_{bar}\sim115$). Also, source 3 (NW) coincides with a
region of distorted isophotes in the optical continuum at P.A. $\sim
-30^{\circ}$.

Figures 3 and 4 reproduce the 2 cm radio continuum emission from the
circumnuclear structures in NGC 1326 and NGC 4314. The source designations
are given in Table 1. The location of these
structures indicate that they are most likely the result of gas
concentrations near (or at) an Inner Lindblad Resonance (ILR) due to the
dynamics driven by the non-axisymmetric potential of the bar (Schwarz
1984). For instance, for a bar angular speed of 36 km s$^{-1}$ kpc$^{-1}$ in
NGC 4314, the ILR could be at a distance of $\sim 450$ pc from the
galactic center and, for a bar angular speed of 60 km s$^{-1}$ kpc$^{-1}$,
the ILR in NGC 1326 could be located between 200 and 400 pc from the
center (see Garc\'{\i}a-Barreto \etal 1991a,b). Gas accumulates near the
resonance, massive stars are then formed and giant \hii regions are
created by the strong photoionization field from these newly formed
massive stars.

\section{Spectral Index of the Radio Continuum Circumnuclear Sources}

We have re-done the maps (see Figs. 1, 3 \& 4), and re-derived the
spectral indices between 20 and 6 cm, and 6 and 2 cm. These indices were
computed from the peak fluxes, determined from gaussian fits made with the
task IMFIT in AIPS. In the case of NGC 1022, the linear resolution was 335
by 135 pc and the average spectral index for the compact nucleus and NE
source between 6 and 2 cm are $\alpha^6_2 \approx +0.12$ and $+0.15$,
respectively. In contrast, we find $\alpha^6_2 \approx +0.9$ for the NW
source. For NGC 1326, the linear resolutions are 295 by 245 pc and the
spectral index is $\alpha^6_2 \approx +0.06$ and for NGC 4314, the linear
resolutions are 195 by 155 pc and the spectral index values are
$\alpha^6_2 \approx +0.2$. The values for all regions are given in Table
1. Only the strongest sources, with flux densities at 2 cm larger than
five times the rms noise values. are considered. Also, our observations do
not resolve individual (compact or giant) \hii regions since their average
size (less than 100 pc) is smaller than our resolution in all three
galaxies.

The spectral index between 20 and 6 cm is negative in all cases, indicating
that the emission in this wavelength range has a non-thermal origin.  
In contrast, the spectral index from 6 to 2 cm is always positive, suggesting
that this emission is dominated by thermal bremsstrahlung.  This is a
consequence of optically thin synchrotron radiation having $\alpha^{20}_6 \sim
-0.8$ (Niklas, Klein \& Wielebinski 1997) and thermal bremsstrahlung having $\alpha$ between
$-$0.1 and +2, depending on the optical depth (\eg, Mezger \& Henderson 1967).
Clearly, the values of $\alpha_2^6$ for all circumnuclear sources in the 
three galaxies are within the range of thermal emission, $-0.1<\alpha_2^6 <+2$
(see Table 1). The total radio continuum flux at these wavelengths is the sum
of synchrotron and free-free emission. However, a detailed decomposition of the
fluxes obtained with similar beams for NGC 1326 by Garc\'\i a-Barreto \etal
(1991b) indicates a very small synchrotron contribution at 2~cm. A similar
exercise with the other sources provides the same result. The synchrotron
emission probably falls more rapidly than the free-free emission rises between
6 and 2~cm, and the positive index values are lower limits (but likely close)
to the true free-free spectral indices. This is clearly seen in Figure 5, that
shows $S_{\nu}$ vs $\nu$ for all the sources listed in Table 1: i) the
emission between 20 and 6 cm falls down less rapidly than typical non-thermal
emission (Niklas, Klein \& Wielebinski 1997), indicating an important thermal contribution at
6 cm, and ii) the extrapolation of the non-thermal component at 2 cm is a
small fraction of the actual emission.

The lower limits to the average electron density values of the emitting
regions are 50 cm$^{-3}$ for NGC 1022 (Garc\'\i a-Barreto \etal 1991c), 8
cm$^{-3}$ for NGC 1326 (Garc\'\i a-Barreto \etal 1991b), and 25 cm$^{-3}$
for NGC 4314 (Garc\'\i a-Barreto \etal 1991a). These values are similar to
the average electron densities, around 10 cm$^{-3}$, reported for the
circumnuclear starburst in NGC 1097 (Hummel, van der Hulst \& Keel 1987),
and to those reported for the giant \hii region W3A, within 1 and 80
cm$^{-3}$ (Kantharia, Anantharamaiah \& Goss 1998)

\section{Discussion}

The spectral indices for the circumnuclear regions in our three barred galaxies
are $\alpha^6_2>-0.1$, and they have similar spatial distributions in the
H$\alpha$, CO and radio continuum emission (Garcia-Barreto et al. 1991a,b,c;
Benedict, Smith \& Kenney 1996). This indicates that massive star formation is
going on in the giant molecular cloud complexes of the observed regions. 

Molecular cloud complexes in our galaxy display highly irregular density and 
velocity distributions, giving the impression of conglomerates of high density 
condensations, or cloud cores, interconnected by a more tenuous inter-core 
medium. These high density condensations seem to be the actual sites of massive
star formation and they host many of the known UC\hii and SUC\hii regions (\eg 
Cesaroni \etal 1999; Kurtz \etal 2000). During the formation phase of an \hii 
region, the radiation field creates an ionization front that evolves within the
density profile of the star forming core. The initial structure and early
expansion phases of the recently formed \hii region, then, are defined by the 
properties of the star-forming core (Franco, Tenorio-Tagle \& Bodenheimer 1989,
1990). Observations of nearby cloud fragments and extinction studies in dark 
clouds indicate density distributions proportional to $r^{-\omega}$, with 
$\omega$ ranging from 1 to 3 and having an average of $\omega \sim 2$ (\eg 
Arquilla \& Goldsmith 1985; Gregorio Hetem, Sanzovo \& Lepine 1988). For 
decreasing density 
gradients with $\omega \geq 1.5$, the ionization front eventually overtakes the
shock front and a large region around the core also becomes ionized. All parts 
of the cloud core are set into motion, sometimes driving internal shocks, and 
instabilities in both the ionization and shock fronts generate clumps and 
finger-like structures (Garc\'{\i}a-Segura \& Franco 1996; Franco \etal 1998;
Williams 1999; Freyer, Hensler \& Yorke 2000). 

The disruptive effects of photoionization and photodissociation first halt the 
star formation activity (Franco, Shore \& Tenorio-Tagle 1994; Diaz-Miller, 
Franco \& Shore 1998) and later, once a large fraction of the parental cloud 
is ionized, accelerate outflows in an expanding giant \hii region. The combined 
action of photoionization and the strong mechanical energy input from stellar 
winds and supernova explosions, then, eventually destroys the molecular cloud 
complex and creates large expanding shells (see reviews by Yorke 1986, 
Tenorio-Tagle \& Bodenheimer 1988, and Bisnovatyi-Kogan \& Silich 1995). Thus,
the non-thermal and thermal radio continuum emission from the circumnuclear
regions of the three barred galaxies can be viewed as resulting from the 
stellar energy input that is destroying the parental clouds. The negative
spectral index between 20 and 6 cm (except for the compact nucleus of NGC 1022)
is likely due to non-thermal emission from SN associated to the star-forming
activity. The positive index between 6 and 2 cm, on the other hand, is likely
due to optically thick thermal emission from the expanding \hii regions.

The spectra of the free-free emission from an optically thick plasma with
various decreasing density stratifications have been calculated by Olnon (1975)
and Panagia \& Felli (1975). The main result of these studies is that, for
unresolved sources, the spectral index is in the range from -0.1 to +2,
depending on the particular functional form of the density gradient and the
optical depth of the region. For a power law of the form $n_e \propto
r^{-\omega}$, the value of the spectral index $\alpha$, S$_{\nu} \propto
\nu^{\alpha}$, depends on the exponent $\omega$ as
(Olnon 1975) 
\beq
\alpha = {(2\omega - 3.1)/(\omega - 0.5)}.
\eeq
For a truncated power-law (which removes the pole $n_e\rightarrow\infty$ at
$r\rightarrow$0) and other functional forms for the density stratification, the
formulae for the spectral index become more complicated. For simplicity, and
following the results from nearby clouds, the power-law case can be used as an
approximation to the plasma density stratification. Thus, the photoionized
circumnuclear clouds detected in the radio continuum can be characterized
by exponents in the range $1.6<{\omega}<2.4$ (the values of $\omega$ for
all sources are listed in Table 1). It is important to emphazise, as the
referee has pointed out, that these giant extragalactic \hii regions are
usually composed of a collection of individual ionized cores emmbedded in
a lower brightness extended region. Thus, the density gradients derived
from the spectral index measurements represent an $average$ density
stratification which may be different from the actual density profiles of
the indiviual ionized cores.

Summarizing, the radio continuum emission from the circumnuclear regions in the
three barred galaxies NGC 1022, 1326 and 4314 show a combination of non-thermal
and thermal features. The fact that all observed sources have similar
properties, a negative index value between 20 and 6 cm and a positive value
between 6 and 2 cm, indicates that the radio emission is mainly due to SN and
\hii regions in massive star forming regions. In particular, the free-free
emission of the unresolved giant \hii regions indicates density gradients
shallower than $\omega \sim 2.5$. This result is similar to the one reported by
Franco \etal (2000) for ultracompact \hii regions, and suggests that
many \hii regions, in our Galaxy and in other galaxies, are optically thick at
radio wavelengths and may provide information of the density stratification of
the emitting plasma. This was already hinted at the reported spectral indices
greater than -0.1 in both giant and ultracompact \hii regions (Wood \&
Churchwell 1989; Kantharia, Anantharamaiah \& Goss 1998), and suggests that
further analysis of the data may provide valuable information about the
properties of star forming regions.

\section*{Acknowledgements}

It is pleasure to thank Stan Kurtz for many stimulating and informative
discussions, and an anonymous referee for useful suggestions that helped
us to improve the contents of the paper. JF acknowledges partial support
by DGAPA-UNAM grant IN130698 and by a R\&D CRAY Research grant. E de la F
wishes to acknowledge financial support from CONACYT-M\'exico grant 124449
and DGEP-UNAM through graduate scholarships.

\clearpage
\newpage

{\center \section*{References}}
\setlength{\parindent}{-1.0\parindent}

Arquilla, R. \& Goldsmith, P. F. 1985, \apj, 297, 436

Benedict, G. F., Smith, B. J. \& Kenney, J. D. P. 1996, \aj, 111, 1861

Bisnovatyi-Kogan, G. S. \& Silich, S. A. 1995, RevModPhys, 67, 661

Cesaroni, R., Felli, M., Jenness, T., Neri, R., Olmi, L., Robberto, M.,
Testi, L. \& Walmsley, C.M. 1999, \aap, 345, 949

Condon, J.J., Cotton, W.D., Greisen, E.W., Yin, Q.F., Perley, R.A., Taylor, G.B. \& Broderick, J.J. 1998, AJ, 115, 1693

D\'{\i}az-Miller, R. I., Franco, J. \& Shore, S. N. 1997, ApJ, 501, 192

Franco, J., Diaz-Miller, R. I., Freyer, T. \& Garc{\'\i}a-Segura, G. 1998,
 in {\it Astrophysics from Antarctica}, ed. G. Novak \& R. H. Landsberg,
ASP Conf. S. 141, 154

Franco, J., Kurtz, S. E., Hofner, P., Garc{\'\i}a-Segura, G. \& Martos,
M. A. 2000, in preparation 

Franco, J., Tenorio-Tagle, G. \&  Bodenheimer 1989, RevMexAA, 18, 65

Franco, J., Tenorio-Tagle, G. \&  Bodenheimer 1990, \apj, 349, 126

Franco, J., Shore, S. N., \& Tenorio-Tagle, G. 1994, \apj, 436, 795

Freyer, T., Hensler, G. \& Yorke, H. 2000, in preparation

Garc{\'\i}a-Segura, G. \& Franco, J. 1996, \apj, 469, 171

Garc{\'\i}a-Barreto, J.A., Downes, D., Combes, F., Gerin, M., Magri, C., 
Carrasco, L., \& Cruz-Gonzalez, I. 1991a, \aap, 244, 257

Garc{\'\i}a-Barreto, J.A., Dettmar, R.-J., Combes, F., Gerin, M., \& 
Koribalski, B. 1991b, RevMexAA, 22, 197

Garc{\'\i}a-Barreto, J.A., Downes, D., Combes, F., Gerin, M., Magri, C., 
Carrasco, L., \& Cruz-Gonzalez, I. 1991c, \aap, 252, 19

Garc{\'\i}a-Barreto, J.A., Franco, J., Carrillo, R., Venegas, S. \& 
Escalante-Ramirez, B. 1996, RevMexAA, 32, 89

Gregorio Hetem, J., Sanzovo, G. \& Lepine, J. 1988, AAS, 76, 347

Hummel, E., van der Hulst, J. \& Keel, W. C. 1987, A\&A, 172, 32

Kantharia, N. G., Anantharamaiah, K. R. \& Goss, W. M. 1998, \apj, 504,
375

Kurtz, S. E. 2000, in {\it Astrophysical Plasmas}, ed. J. Arthur, N.
 Brickhouse \& J. Franco, RevMexAA (Conf. Ser.), in press 

Kurtz, S., Cesaroni, R., Churchwell, E., Hofner, P, \& Walmsley, C.M.
2000, in {\it Protostars \& Planets IV}, ed. V. Mannings, A. Boss \& S.
Russell, in press

Mezger, P.G. \& Henderson, A. P. 1967, \apj, 147, 471

Mezger, P. G., Schraml, J. \& Terzian, Y. 1967, \apj, 150, 807
 
Niklas, S., Klein, U., \& Wielebinski, R. 1997, A\&A, 322, 19
 
Panagia, N. \& Felli, M. 1975, A\&A, 39, 1
 
Olnon, F. M. 1975, A\&A, 39, 217

%Reynolds, S. P. 1986, \apj, 304, 713

Schwarz, M. P. 1984, \mnras, 209, 93
 
Simon, M., Felli, M., Casser, L., Fischer, J., \& Massi, M. 1983, ApJ,
266, 623

Tenorio-Tagle, G. \& Bodenheimer, P. 1988, ARAA,  26, 145

Tully, R. B. 1988, in {\it Nearby Galaxy Catalog}, (Cambridge:Cambridge
Univ. Press)

Williams, R.J.R. 1999, MNRAS, 310, 789

Wood, D. O. S. \& Churchwell, E. 1989, \apjs, 69, 831

%Wright, A. E. \& Barlow, M. J. 1975, \mnras, 170, 41  

Yorke, H. W. 1986, ARAA, 24, 46

\normalsize 

\clearpage
%\newpage

%\section*{Tables}

%Table 1

\small 
\begin{table}
\caption[ ]{
Radio Continuum Spectral Indices}
\begin{flushleft}
\begin{tabular}{cclccc}
\hline
\\
Galaxy & Source  & Description 	    &\ $\alpha^{20}_6$		& $\alpha^6_2$  & $\omega^a$  
\cr
\hline
\\ 
NGC 1022 & NGC 1022:FGF 1$^b$ & Compact Nucleus & -0.83$\pm0.03$ & +0.12$\pm0.03$ & 1.62 
\cr
\\
 & NGC 1022:FGF 2 & North East source & -0.6$\pm0.1$ & +0.15$\pm0.05$ &  1.63  
\cr
\\
 & NGC 1022:FGF 3 & North West source & -0.9$\pm0.3$  & +0.9$\pm0.3$  & 2.41 
\cr
\\ 
NGC 1326 & NGC 1326:FGF 1 &  Western source  & -0.67$\pm0.03$   & +0.06$\pm0.05$  & 1.58  
\cr
\\
NGC 4314 & NGC 4314:FGF 1 &  North source & -0.83$^{+0.1}_{-0.04}$   &  +0.26$^{+0.03}_{-0.1}$ & 1.70 
 \cr
\\
         & NGC 4314:FGF 2 & North East source & -0.82$\pm0.2$  &  +0.23$\pm0.2$   & 1.68 
 \cr
\\
         & NGC 4314:FGF 3 & South East source & -0.63$\pm0.05$  &  +0.18$\pm0.1$   & 1.65 
 \cr
\\
\hline  
\end{tabular}
\end{flushleft}
$^a$ $\omega$ is the exponent of the density distribution, n$_e\propto~$r$^{-\omega}$ (see text).

$^b$ Source also known as NWSS J023832-064039 (Condon et al. 1998)
\end{table}

\normalsize 
\clearpage

\newpage

\figcaption{Radio continuum emission at 2 cm (contours) superimposed on the 
H$\alpha$ emission (grey scale) of NGC 1022 (from Garc\'{\i}a-Barreto \etal 
1991c; 1996). Grey scale: 2.6 to 170$~\times 10^{-15}$ ergs cm$^{-2}$ seg$^{-1}$/beam. Contours: 2.5, 3, 4, 6, 7, 11, 15, 17, 19, 23, 27, 
and 31 times 150 ${\mu}$Jy/beam (rms noise is 150 ${\mu}$Jy/beam).
Restoring beam FWHM $\sim 3''.7 \times 1''.5$ at P.A. $\sim -11^{\circ}$, corresponding to a linear scale of 335 pc by 135 pc.
 \label{fig1}}

\figcaption{Optical continuum emission from the central region of NGC 1022 
(contours) superimposed on the radio continuum emission at 20 cm (grey scale) 
(images taken from Garc\'{\i}a-Barreto \etal 1991c; 1996). Grey scale: 0.3 to 10 mJy/beam. Contours (in arbitrary brigtness units of 10): 9, 9.5, 10, 10.5, 11, 11.5, 12, 12.5, 13, 14, 15, 16, 17, 18, 19, 20, 21, 22, 23, 24, 25, 27.5, 30, 40, and 50. \label{fig2}}

\figcaption{Radio continuum emission at 2 cm (contours and grey scale) of NGC 
1326 taken from Garc\'{\i}a-Barreto \etal (1991b). Grey scale: ~0.4~to~2.2~mJy/beam. Contours: -1, 1.5, 3, 4, 5, 7, and 9 times 250
${\mu}$Jy/beam (rms noise is 250 ${\mu}$Jy/beam). Restoring beam FWHM
$\sim 3''.6 \times 3''$ at P.A. $\sim 30^{\circ}$ corresponding to a linear scale of 295 pc by 245 pc. \label{fig3}}

\figcaption{Radio continuum emission at 2 cm (contours and grey scale) of NGC 
4314 taken from Garc\'{\i}a-Barreto \etal (1991a). Grey scale: 45 to 320 ${\mu}$Jy/beam. Contours: -2, 2, 2.5, 3, 4, 5, 5.5, 6, 6.5, and 7 times 45
${\mu}$Jy/beam (rms noise is 45 ${\mu}$Jy/beam). Restoring beam FWHM $\sim
3''.9 \times 3''.1$ at P.A. $\sim 90^{\circ}$ corresponding to a linear scale of 195 pc by 155 pc.
\label{fig4}}

\figcaption{$S_{\nu}$ vs $\nu$ for all the sources listed in Table
1. For clarity, the sources are grouped by galaxy (solid, dotted and
dashed lines correspond to sources 1, 2 and 3, respectively), and each
curve has been multiplied by the factor indicated in parenthesis.
  \label{fig5}}
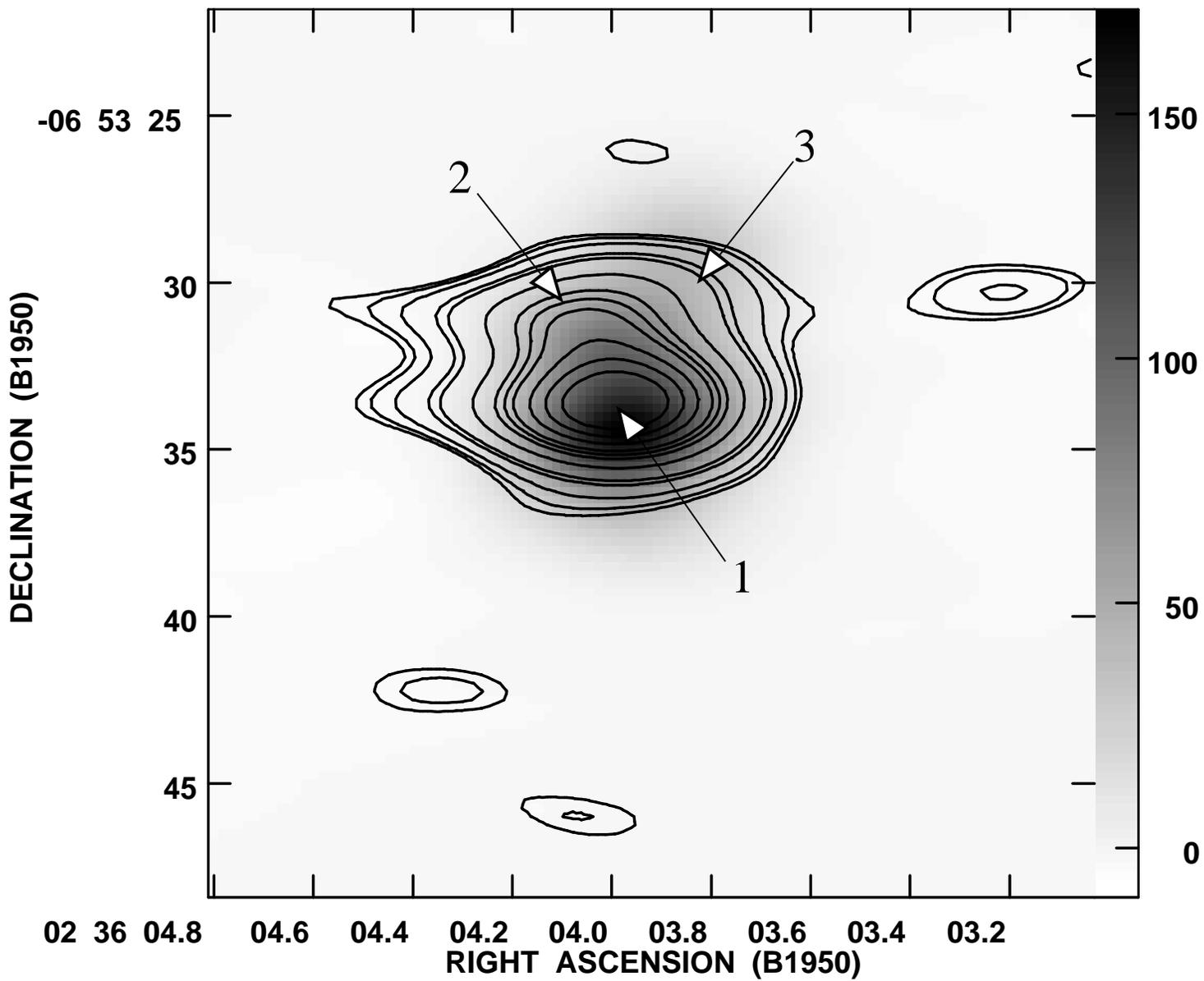
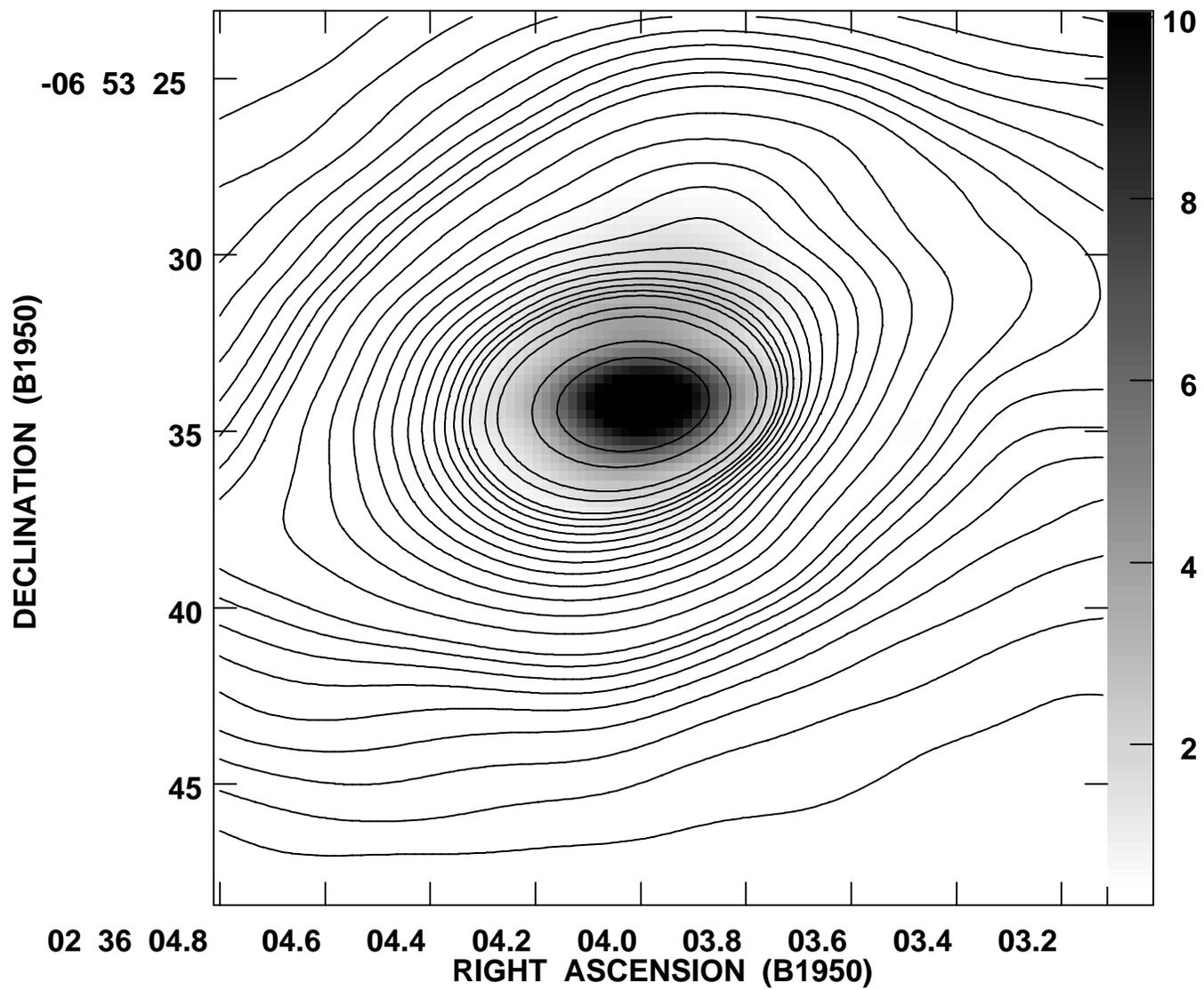
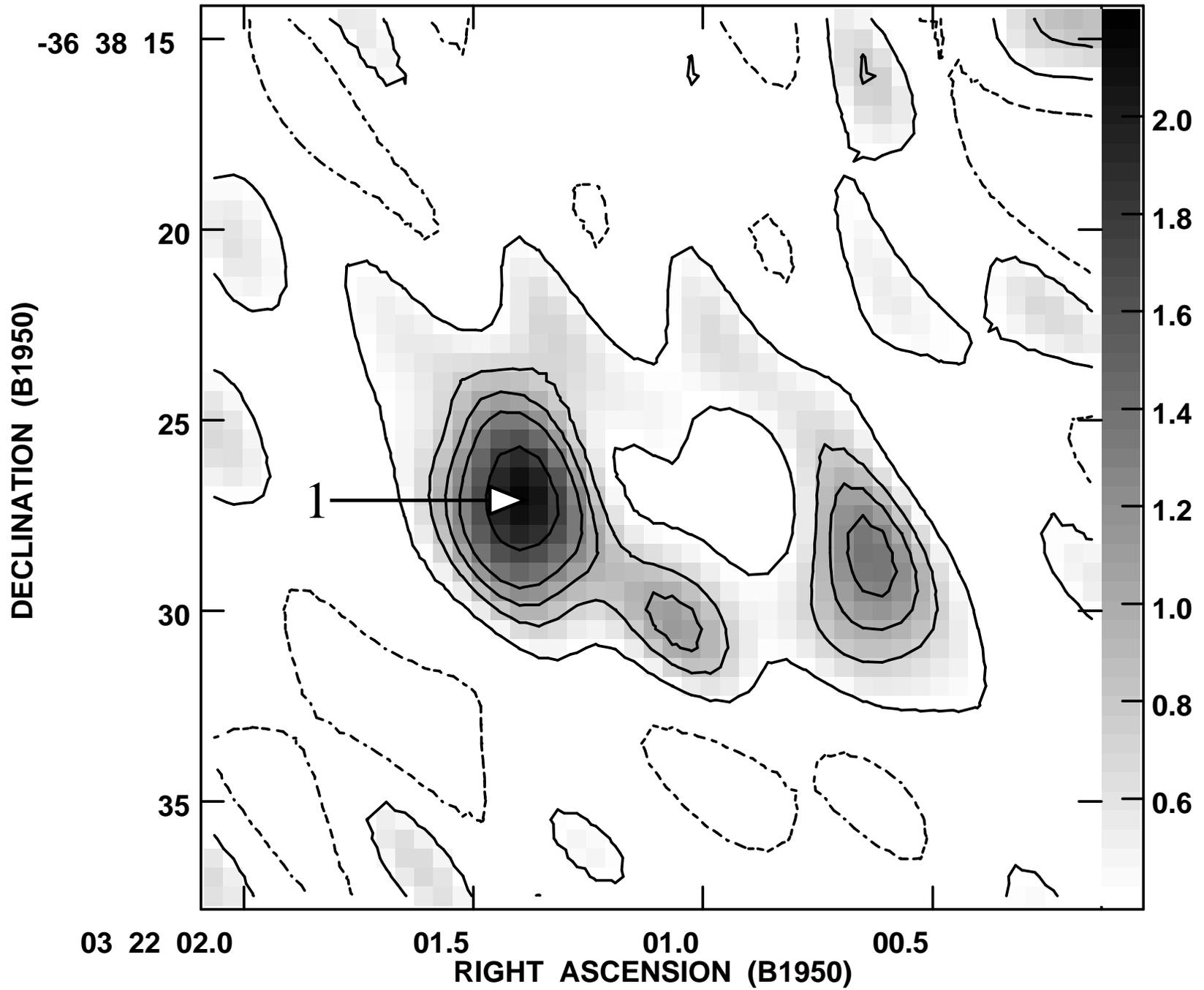
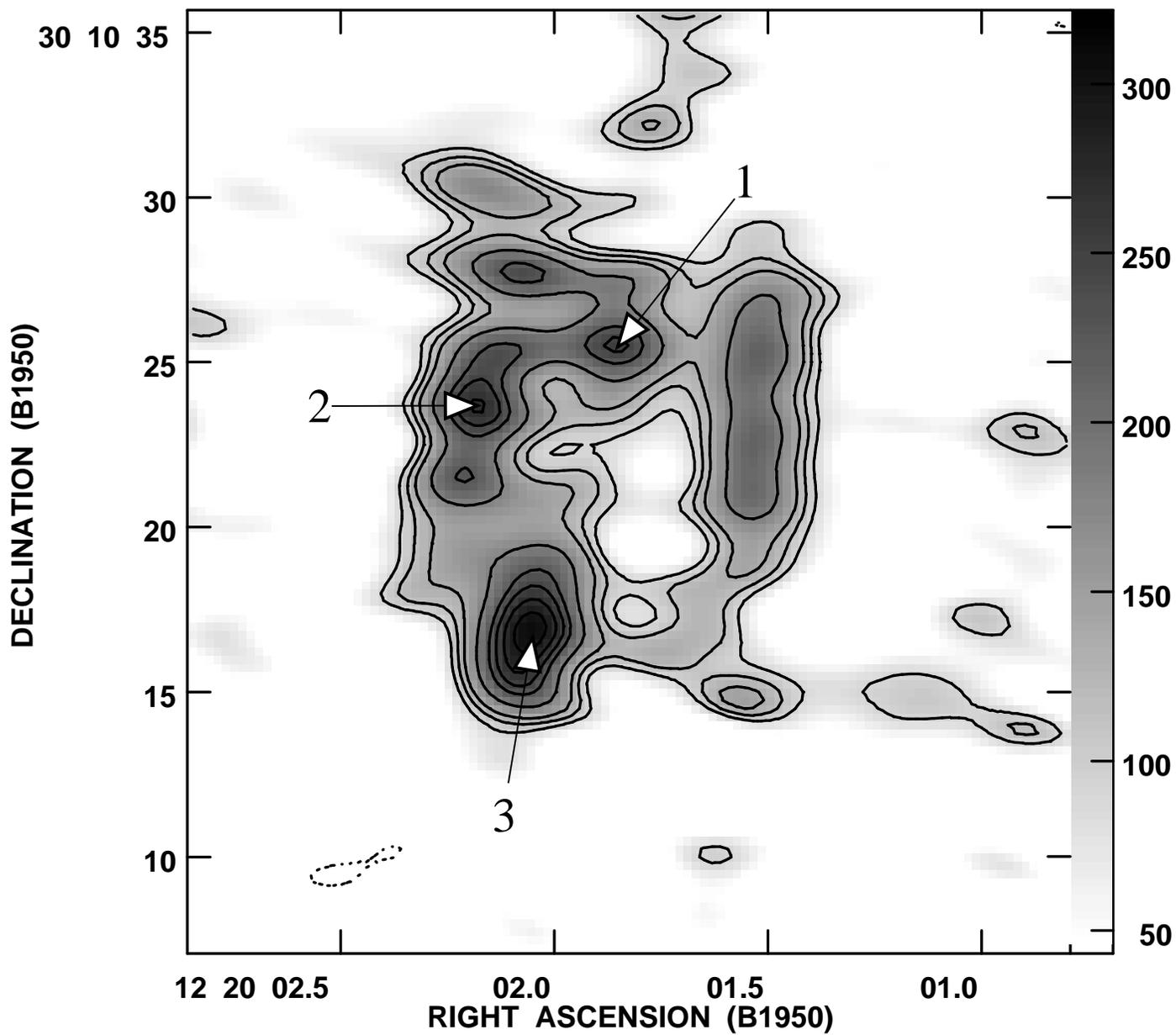
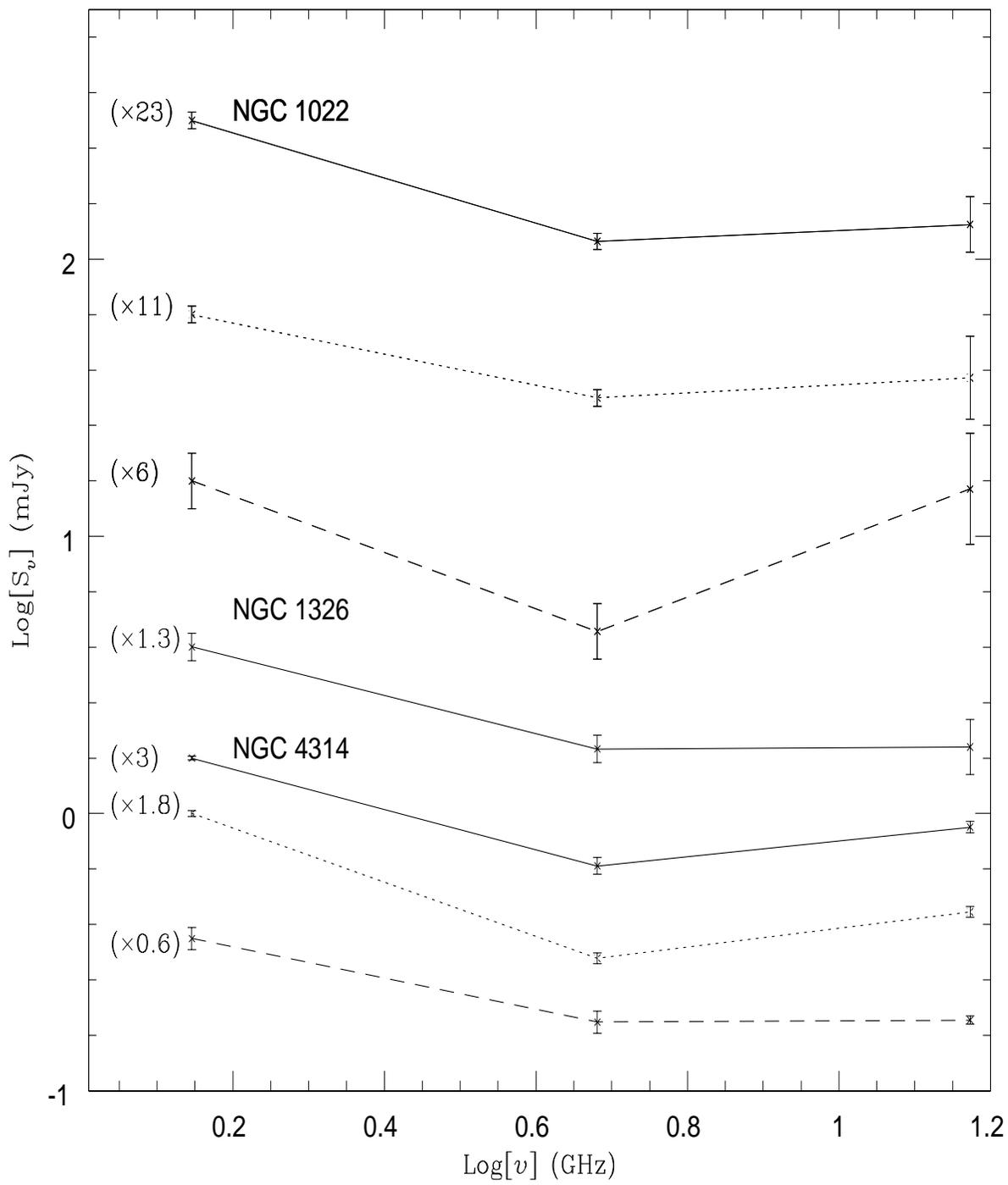
\end{document}